\documentclass[twocolumn,showpacs,amsmath,amssymb]{revtex4}
\usepackage{graphicx}
\usepackage{dcolumn}
\usepackage{bm}

\newcommand{\be}{\begin{equation}}
\newcommand{\ee}{\end{equation}}
\newcommand{\bea}{\begin{eqnarray}}
\newcommand{\eea}{\end{eqnarray}}
\begin{document}

\title{Dimerized phase and transitions in a spatially anisotropic square
  lattice antiferromagnet}

\author{Oleg A. Starykh$^1$} \author{Leon Balents$^2$}
\affiliation{$^1$Department of Physics and Astronomy, Hofstra
  University, Hempstead, NY 11549 \\ $^2$Department of Physics,
  University of California, Santa Barbara, CA 93106-4030 }

\begin{abstract}
  We investigate the spatially anisotropic square lattice quantum
  antiferromagnet.  The model describes isotropic spin-$1/2$ Heisenberg
  chains (exchange constant $J$) coupled antiferromagnetically in the
  transverse ($J_\perp$) and diagonal ($J_\times$), with respect to the
  chain, directions.  Classically, the model admits two ordered ground
  states -- with antiferromagnetic and ferromagnetic inter-chain spin
  correlations -- separated by a first order phase transition at
  $J_\perp=2J_\times$. We show that in the quantum model this transition
  splits into two, revealing an intermediate quantum-disordered columnar
  dimer phase, both in two dimensions and in a simpler two-leg ladder
  version. We describe quantum-critical points separating this
  spontaneously dimerized phase from classical ones.
\end{abstract}
\pacs{}

\maketitle 

An interplay between geometric frustration and quantum fluctuations is
at the heart of intensive current investigations into the nature of
possible $SU(2)$-invariant Mott insulators and quantum phase
transitions.  The absence of any ``natural'' small
parameter, however, makes a traditional perturbative approach difficult.
Aside from
direct numerical attacks involving exact diagonalization and quantum
Monte-Carlo techniques, this feature forces one to extend the parameter
space of frustrated magnetic models and explore various ``corners'' of
the resulting phase diagram in the hope of gaining new insights into the
physically relevant region of system's parameters.
In this paper, we explore a simple {\sl spatially anisotropic}
frustrated spin model which allows the application of powerful
analytical methods borrowed from one dimension.  

The model we consider -- the spatially anisotropic square lattice
antiferromagnet -- is a quasi-one-dimensional generalization of the
well-studied square lattice antiferromagnet with frustrating
antiferromagnetic exchange along diagonals, also known as the $J_1-J_2$
model.
It describes a collection of antiferromagnetic spin$-S$ chains with
exchange constant $J$ running along horizontal (chain) direction.  The
chains are interacting with their nearest neighbors via weak
antiferromagnetic spin exchange in the transverse ($J_\perp$) and
diagonal ($J_\times$), with respect to the chain, directions.  Recently
this model was investigated by Nersesyan and Tsvelik (NT) \cite{NT} who
predicted a novel RVB-like phase with deconfined massive spinons at the
special ratio of microscopic exchange constants: $J_\perp=2J_\times \ll
J$.  We shall comment on this work and subsequent numerical
investigations \cite{numerics,sindzingre} at the end of our paper.
Let integer index $n$ numerate sites along the chain while index $m$ numerates
chains. Then the Hamiltonian is $H=\sum_m H_{m}^{(0)} + V$, where
$H_{m}^{(0)}$ is the standard Heisenberg Hamiltonian of the $m$-th chain and the
interchain interaction reads
\begin{eqnarray}
&& \hspace{-0.3in}V=\sum_{n,m}\Big\{ J_\perp S^a_{n,m} S^a_{n,m+1} + J_\times  \sum_{q=\pm
  1}S^a_{n,m} S^a_{n+q,m+1}\Big\} .
\label{V-full}
\end{eqnarray} 
Here $a=x,y,z$ is the vector index over which implicit summation is
implied.  The classical ($S=\infty$) phase diagram is simple: for
$J_\perp > 2 J_\times$ ordering of the spins in the transverse to chain
direction is antiferromagnetic (AFM phase), whereas for $J_\perp < 2
J_\times$ it is ferromagnetic (FM phase); spins order
antiferromagnetically along chains. In the isotropic $J_1-J_2$ model
these orderings correspond to the N\'eel and four-sublattice phases,
respectively.  The line $J_\perp = 2 J_\times$ describes a first order
transition between these two phases. Along this line the interchain
interaction can be written in a more suggestive form
\begin{equation} 
V_\times=
\sum_{n,m} J_\times (S^a_{n,m} + S^a_{n+1,m})(S^a_{n,m+1}
+S^a_{n+1,m+1}).
\label{V-times}
\end{equation} 
Our goal is to understand the phase diagram of the quantum $S=1/2$
model in the limit of weak interchain couplings $J_\perp, J_\times \ll
J$. Taking the continuum limit along the chain direction ($x=na$, $a$ is
the lattice spacing), the spin operator is decomposed as 
\begin{equation}
S^a_{n,m}\rightarrow a\Big(J^a_{m,R}(x) + J^a_{m,L}(x) + (-1)^n
N^a_m(x)\Big)
\label{spin}
\end{equation}
in terms of chiral components $J^a_{m,R/L}$ of the
uniform spin magnetization of the $m$-th chain,
and the staggered spin magnetization $N^a_m$ (scaling dimension $1/2$).
The uniform spin magnetization $J^a_m=J^a_{m,R} + J^a_{m,L}$ is the conserved spin current (scaling dimension $1$)
and a generator of  $O(3)$ spin rotations. 
The two fields in (\ref{spin}) are connected via the following 
operator product expansion (OPE) 
\begin{equation}
J^a_{R/L}(x,\tau) N^b(x',\tau') = 
\frac{\mp i\delta^{ab} \epsilon(x',\tau') + i\epsilon^{abc}
  N^c(x',\tau')}{4\pi [v(\tau-\tau')\mp i(x-x')]} 
\label{ope}
\end{equation}
Here the upper/lower signs on the right hand side apply for the
right/left (R/L) moving currents, respectively, and $v=\pi Ja/2$ is the
spin velocity.  Of major importance is the appearance of the staggered
{\em dimerization} operator $\epsilon$ (scaling dimension $1/2$).  It is
the continuum limit of the lattice operator involving the scalar product
of two neighboring spins, $\epsilon_m(x) \sim (-1)^n S^a_{n,m}
S^a_{n+1,m}$.

The spin algebra encoded in (\ref{ope}) and well-known OPE
of the chiral components of the spin current \cite{tsvelik-book} follow
from the fact that the Hamiltonian density of the isolated Heisenberg chain
is described by the $SU_1(2)$ WZW model perturbed by a marginally
irrelevant backscattering term $(g_{\text{bs}}=O(J))$
\begin{equation}
{\cal H}^{(0)}_m=\frac{2\pi v}{3}( J^a_{m,R} J^a_{m,R}+ J^a_{m,L}
J^a_{m,L}) - g_{\text{bs}}J^a_{m,R}J^a_{m,L} .
\label{wzw}
\end{equation}
The c-functions appearing in the OPE above are just the Green's functions of right- and left-moving fermions
which are governed by the Hamiltonian (\ref{wzw}) with $g_{\text{bs}}=0$.

Having exposed the structure of the unperturbed  theory, we now turn to the inter-chain interaction (\ref{V-full}).
Its continuum low-energy Hamiltonian density contains all terms allowed by the symmetries of the lattice
model (these include reflection with respect to transverse direction and translation by one lattice constant)
\begin{eqnarray}
{\cal V}&=&\sum_m g_1 N^a_m N^a_{m+1} +g_2 J^a_m J^a_{m+1} \nonumber\\
&& + g_3 a^2 \partial_x N^a_m \partial_x N^a_{m+1} + g_4 \epsilon_{m}\epsilon_{m+1}
\label{V-full-cont}
\end{eqnarray}
Here $\partial_x$ denotes derivative with respect to $x$ and the bare
coupling constants follow from substituting (\ref{spin}) in
(\ref{V-full})
\begin{equation}
g_1=(J_\perp - 2 J_\times)a, g_2=(J_\perp + 2 J_\times)a,
g_3=\frac{J_\perp a}{2},   g_4=0.
\label{couplings}
\end{equation} 
The scaling dimensions of the first, second, and third interaction terms
in (\ref{V-full-cont}) are $1$, $2$, and $3$, respectively. The last
term (scaling dimension $1$) is included here because it respects the
symmetries of the lattice model (\ref{V-full}). Although its bare
coupling constant is zero, quantum fluctuations can (and will) generate
some finite $g_4$.

At this level the phase diagrams of the quantum model,
(\ref{V-full-cont}) with $g_4=0$, and classical one, (\ref{V-full}) with
$S=\infty$, are identical.  As long as $J_\perp \neq 2J_\times$,
(\ref{V-full-cont}) is dominated by the strongly relevant interaction of
staggered magnetizations which drives the system into one of the
classically ordered phases: AFM for $g_1 >0$ and FM for $g_1<0$.

However, the transitional region $J_\perp \approx 2J_\times$ requires
a closer look because there the amplitudes of the both relevant terms are
zero.  To begin, let us fine-tune to the point $g_1=0$. On the lattice
this corresponds to (\ref{V-times}), whose continuum limit is given by
(\ref{V-full-cont}) with $g_1=g_4=0$,
\begin{equation}
{\cal V}_\times=\sum_m g_2 J^a_m J^a_{m+1} + g_3 a^2 \partial_x N^a_m
\partial_x N^a_{m+1} .
\label{V-times-cont}
\end{equation}
The situation is now apparently controlled by the marginally relevant
($g_2 >0$) current-current interaction which seems to suggest that the
strongly irrelevant $g_3$-term does not play any role.  Recall that the
(Kac-Moody) algebra of spin currents $J^a_{R/L}$ is closed in the sense
that OPE of two like currents produces another current of the same
chirality \cite{tsvelik-book}.  Thus ${\cal V}_\times$ with $g_3=0$ will
``reproduce'' itself in every order of the expansion in powers of the
coupling constant $g_2$. This peculiar behavior is destroyed by $g_3\neq
0$ (or any other equally or even more irrelevant term).  To see so, we
expand the partition function to second order in ${\cal V}_\times$.
As follows from the
discussion above, one should concentrate on the {\em cross-term}
($\propto g_2 g_3$)
\begin{equation}
\int_{x,\tau} \int_{x',\tau'} J^a_m(x',\tau') \partial_x N^b_m(x,\tau) 
J^a_{m+1}(x',\tau') \partial_x N^b_{m+1}(x,\tau).
\nonumber
\end{equation}
Applying the OPE (\ref{ope}) to the same-chain operators at nearby
points, differentiating the c-function pre-factors with respect to $x$
and performing elementary integration over the relative coordinates, 
one obtains a {\em relevant} correction to the Hamiltonian of the same
form as (\ref{V-full-cont}).
The coefficients in (\ref{couplings}) are thereby replaced by
\begin{equation}
g_1=(J_\perp -2J_\times + \frac{2J_\times^2}{\pi^2 J})a,
~g_4=-\frac{3J_\times^2 a}{\pi^2 J}, 
\label{couplings-ren}
\end{equation}
while two other couplings $g_{2,3}$ remain unchanged.  Exactly the same
result (\ref{couplings-ren}) can be obtained by performing the standard
renormalization group (RG) analysis of (\ref{wzw}) and
(\ref{V-full-cont}). There one obtains five coupled non-linear
differential equations for $g_{1-4}$ and $g_{\text{bs}}$.  In the
transitional region, corresponding to $g_{1,4}= O(J_\times^2/J) \ll
J_\times$, RG equations decouple and admit a simple analytical solution,
which reproduces (\ref{couplings-ren}), for $\ell$ of the order $1$
(here $\ell$ parametrizes the short-distance cutoff:
$a_\ell=a_0e^\ell$).  Our derivation of the renormalized coupling
constants is very similar in spirit to the recent calculation of
second-order corrections to coupling constants of two relevant and
competing interactions in the extended Hubbard model \cite{akira}.  

What are the phases of ${\cal V}$ with relevant couplings given by (\ref{couplings-ren})?
We begin our analysis by studying two coupled chains, that is, frustrated
spin ladder ($m=1,2$ in (\ref{V-full-cont})). Hamiltonian density of the ladder 
${\cal H}_{\text{ladder}}={\cal H}^{(0)}_1 + {\cal H}^{(0)}_2 + {\cal V}$ can be recast as
a theory of four massive real (Majorana) fermions, see Ch. 21 of \cite{tsvelik-book},
\begin{eqnarray}
{\cal H}_{\text{ladder}}&=&\sum_{a=1}^4 (-\frac{iv}{2} \xi^a_R \partial_x \xi^a_R + \frac{iv}{2}
\xi^a_L \partial_x \xi^a_L)\nonumber\\
 &&-im_t \sum_{a=1}^3 \xi^a_R \xi^a_L -im_s \xi^4_R \xi^4_L + \delta {\cal H}_{\text{marg}}
 \label{ladder-majorana}
\end{eqnarray}
where $\delta {\cal H}_{\text{marg}}$, originating from $g_{\text{bs}}$
in (\ref{wzw}) and $g_2$ in (\ref{V-full-cont}), describes residual {\em
  marginal} interactions between Majorana fermions
\begin{eqnarray}
&&  \hspace{-0.3in} \delta {\cal H}_{\text{marg}}=\frac{g_-}{4} \Big(\sum_{a=1}^3 \xi^a_R \xi^a_L\Big)^2
 - \frac{g_+}{2} \sum_{a=1}^3 \xi^a_R \xi^a_L \xi^4_R \xi^4_L,
\label{ladder-marg}
 \end{eqnarray}
 with $g_\pm = g_2 \pm g_{\rm bs}$.  Ignoring $\delta {\cal
   H}_{\text{marg}}$, the first three Majorana fermions form a triplet
 with the mass
 \begin{equation}
 m_t=\frac{\lambda^2(g_1 - g_4)}{2\pi a}=\frac{\lambda^2}{2\pi}(J_\perp-2J_\times + \frac{5J_\times^2}{\pi^2 J}),
 \label{m-t}
 \end{equation}
 whereas the fourth fermion $\xi^4_{R/L}$ is a singlet with
 \begin{equation}
 m_s=-\frac{\lambda^2(3g_1+g_4)}{2\pi
   a}=-\frac{3\lambda^2}{2\pi}(J_\perp-2J_\times +
 \frac{J_\times^2}{\pi^2 J}).
 \label{m-s}
 \end{equation}
 Both masses receive {\em finite} logarithmic corrections from $\delta
 {\cal H}_{\text{marg}}$ which we omit in the following.  Here $\lambda
 \sim 1$ is the expectation value of the charge operator which
 multiplies abelian bosonization expressions for the staggered
 magnetization and dimerization operators \cite{shelton-ladder}
 \begin{eqnarray}
 {\bf N}_m &= &\frac{\lambda}{\pi a}(\cos\sqrt{2\pi}\theta_m, \sin\sqrt{2\pi}\theta_m,-\sin\sqrt{2\pi}\varphi_m),\nonumber
 \\
 \epsilon_m&=&\frac{\lambda}{\pi a}\cos\sqrt{2\pi}\varphi_m.
 \label{bosonized}
\end{eqnarray}
The Hamiltonian (\ref{ladder-majorana}) predicts {\em two} transitions
\cite{nt-ladder} which happen when one
of the renormalized masses becomes zero.\\
({\em i}) $m_t=0$: at this point triplet excitations become gapless.
Integrating out massive fermion $\xi^4$, we find that ${\cal
  H}_{\text{ladder}}$ is the Hamiltonian of the $SU_2(2)$ WZW model with
central charge $c=3/2$. The nature of the transition is determined by
the sign of remaining marginal coupling $g_2 - g_{\text{bs}}$.  Its
positive initial value implies marginally relevant flow to the strong
coupling, and hence, first order phase transition.  For the negative
initial value one obtains marginally irrelevant flow of the coupling
constant to zero, and hence, second order transition.\\
({\em ii}) $m_s=0$: singlet excitations become massless. This is a
continuous $Z_2$ (Ising) transition.
 
Properties of the various phases are most conveniently understood from the
bosonized form of the {\em relevant} part of interchain interaction
(\ref{V-full-cont}), ${\cal V}_{\text{rel}}$, which corresponds to mass
terms in (\ref{ladder-majorana}).  In terms of symmetric and
antisymmetric combinations of bosonic fields on two chains,
$\varphi_\pm=(\varphi_1 \pm \varphi_2)/\sqrt{2}$ (and similarly for
$\theta_\pm$) it reads
 \begin{eqnarray}
 {\cal V}_{\text{rel}}&=&\frac{\lambda^2}{2(\pi a)^2}\Big( 2g_1 \cos\sqrt{4\pi}\theta_{-} + (g_1 + g_4)\cos
 \sqrt{4\pi}\varphi_{-} \nonumber\\
 &&- (g_1 -g_4)\cos\sqrt{4\pi}\varphi_{+}\Big)
 \label{eff-potential}
 \end{eqnarray}
 Depending on the ratio $g_1/g_4$ four phases are possible:\\
 (a) $g_1 \rightarrow -\infty$ is the Haldane (effective spin-$1$ chain)
 phase; $m_t <0$, $m_s >0$ and $\varphi_{+}=\sqrt{\pi}/2,
 ~\theta_{-}=0$. Inter-chain spin correlations are ferromagnetic.
 This is the ladder analog of the classical FM phase.\\
 (b) $g_1 \rightarrow +\infty$ is the rung-singlet phase; $m_t > 0$,
 $m_s < 0$ and $\varphi_{+}=0,~\theta_{-}=\sqrt{\pi}/2$. Here
 inter-chain spin correlations are antiferromagnetic,
 which corresponds to the classical AFM phase.\\
 (c) $g_4 \rightarrow -\infty$ is the columnar dimer phase (DC); $m_{s,t} >
 0$ and
 $\varphi_{+}=0,~\varphi_{-}=0$.\\
 (d) $g_4 \rightarrow +\infty$ is the staggered dimer phase (DS); $m_{s,t}
 < 0$ and
 $\varphi_{+}=\sqrt{\pi}/2,~\varphi_{-}=\sqrt{\pi}/2$.\\
 The last two phases are spontaneously dimerized and do not have
 classical prototypes.  Using (\ref{bosonized}) we find that both
 dimerized phases break translational symmetry (along the chain
 direction) but their ordering patterns are different. In the DC phase
 $\langle \epsilon_1\rangle = \langle \epsilon_2\rangle$, which
 describes {\em columnar dimer} long-range order.  In the DS phase
 $\langle \epsilon_1\rangle = -\langle \epsilon_2\rangle$, which makes
 it the {\em staggered dimer} state. 
Observe that
 microscopic couplings (\ref{couplings-ren}) choose {\em columnar dimer}
 as the intermediate phase between the Haldane and rung-singlet phases
 of the frustrated ladder. 
 
\begin{figure}
\center
\includegraphics[width=0.9\columnwidth]{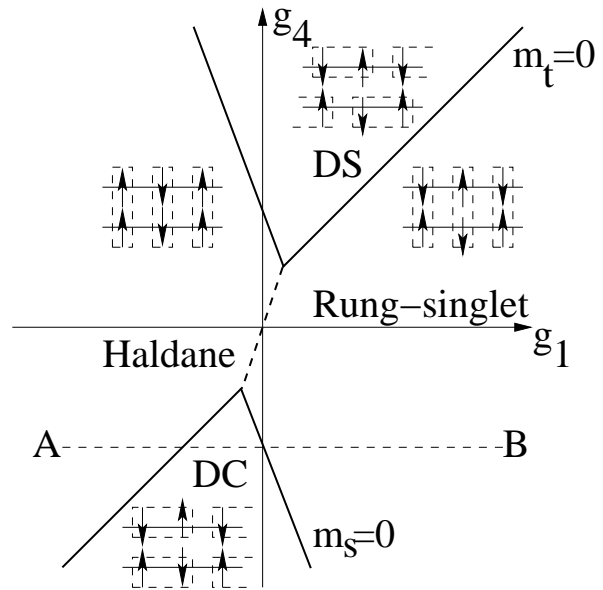}
\caption{Phase diagram of the frustrated ladder. 
Line AB marks the parameters corresponding to the lattice model.}
\label{cfm-ladder}
\end{figure}

 This feature was missed by \cite{allen-essler-ners,solyom} who
 described $J_\perp=2J_\times$ point of the lattice ladder by
 (\ref{V-times-cont}) with $g_3=0$.  Our analysis shows that
 multicritical point where chains are coupled only by the marginal
 current-current interaction requires additional fine-tuning of {\sl
   both} $g_1,g_4$ (equally, of $m_s,m_t$) to zero.  A more detailed understanding
 can be gained by considering the model $g_2 >0$ and {\sl exponentially}
 small relevant couplings $g_{1,4}$ (more specifically $1/|\ln g_{1,4}|
 \ll g_2$) -- though this is not realized by the lattice model.  In this
 case the NT analysis\cite{NT} can be taken over, treating $g_{1,4}$ as
 small perturbations.  A semiclassical analysis for $g_{1,4}=0$ shows
 that the NT model has a manifold of ground states in which minima of
 type (a) and (b) above -- i.e. Haldane and rung-singlet states -- are
 degenerate.  Perturbing with (\ref{eff-potential}) above {\sl splits}
 the {\sl energy densities} of these two degenerate states, so that the
 $g_4=3g_1$ becomes a {\sl first order transition} line 
 (dashed line through the origin in Fig.~\ref{cfm-ladder}).  
 Furthermore,
 the {\sl spinons} of the NT phase correspond to domain walls between
 these two coexisting states.  Remembering that the finite length spin
 chains in the Haldane phase are characterized by spin-$1/2$ end states,
 the spinons become physically transparent as mobile versions of these.
 Away from the first order line, such spinons are therefore {\sl
   confined} (i.e. cost an energy linear in system size), due to the
 energy cost to create a domain of the disfavored phase.  Thus although
 it does not occur in the lattice model studied here, the NT state and
 its spinons can exist as a first order line in a generalized ladder
 model.

Having understood the ladder, we tackle the full two-dimensional
problem.  We treat the relevant inter-chain interactions $g_{1,4}$
(taking $g_{2,3}=0$) using a self-consistent mean-field approximation\cite{schulz,essler}.
Assuming symmetry breaking along $z$-direction,
\begin{equation}
{\cal V}=-z (|g_1| \langle N^z_m\rangle N^z_m + |g_4| \langle \epsilon_m\rangle \epsilon_m)
\label{V-mf}
\end{equation}
where we used $\langle N^z_{m+1}\rangle = \pm \langle N^z_m\rangle$ for
$g_1$ negative (positive), and $z=2$ is the number of neighboring
chains. Using (\ref{bosonized}) and bosonized form of ${\cal H}^{(0)}_m$
(with $g_{\text{bs}}=0$), rescaling euclidian time $\tau=y/v$ and
introducing dimensionless euclidian distance $r=(x,y)/a$, we arrive at
the following single-chain sine-Gordon action
\begin{equation}
S_{\text{mf}}=\int d^2r \Big(\frac{1}{2} (\nabla_r \varphi)^2 - p
\sin\sqrt{2\pi}\varphi - q \cos\sqrt{2\pi}\varphi\Big) 
\label{Smf}
\end{equation}
where dimensionless parameters
\begin{equation}
p\equiv \frac{z\lambda^2 |g_1|}{\pi^2v}\langle \sin\sqrt{2\pi}\varphi\rangle , 
~~q\equiv \frac{z\lambda^2 |g_4|}{\pi^2v}\langle \cos\sqrt{2\pi}\varphi\rangle
\label{p-q}
\end{equation}
are effective staggered magnetization and dimerization fields,
respectively. These averages are found by differentiating the free
energy density $F(p,q)=-\ln Z/V, ~Z={\text {Tr}}(e^{-S_{\text{mf}}})$
with respect to $p$ and $q$, respectively.  Clearly from (\ref{Smf}), the mean-field free
energy can depend on $p,q$ only through
$\kappa=\sqrt{p^2 + q^2}$.
One can therefore take $F(p,q)=F(0,\kappa)$ and take advantage of exact
results for the standard sine-Gordon action\cite{LZ}. 
In particular $dF/d\kappa= - c_1\kappa^{1/3}$, where the numerical
constant $c_1$ follows from equations (10-14) of ref.\cite{LZ}
\begin{equation}
c_1=\frac{\tan(\pi/6)}{3} \Big(\frac{2
  \Gamma(1/6)}{\sqrt{\pi}\Gamma(2/3)}\Big)^2 \Big(\frac{\pi
  \Gamma(3/4)}{2\Gamma(1/4)}\Big)^{4/3}. 
\end{equation}
Thus our self-consistent equations 
\begin{equation}
\langle \sin\sqrt{2\pi} \varphi\rangle = \frac{c_1 p}{\kappa^{1/3}}  , 
~~\langle \cos\sqrt{2\pi} \varphi\rangle = \frac{c_1 q}{\kappa^{1/3}}
\label{m-f}
\end{equation}
become simple algebraic ones. They predict two transitions, at $g_1=\pm|g_4|$, 
where both order parameters ($N^z_m,\epsilon_m\sim p,q$) are
non-zero.  
This makes them {\em first-order} transitions -- the first order nature
being likely an artifact of the mean-field approximation.
\begin{figure}
\center
\includegraphics[width=0.8\columnwidth]{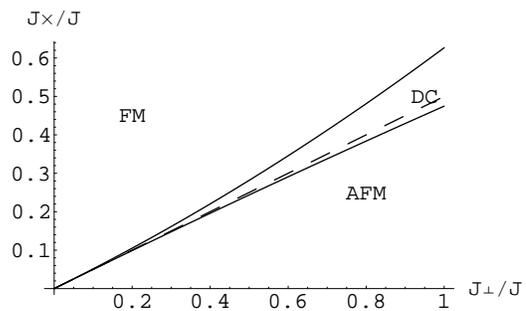}
\caption{Phase diagram of the two-dimensional model. 
Solid (dashed) lines denote phase boundaries of the quantum (classical) model.}
\label{cfm-2dphase}
\end{figure}
These two transitions separate three phases, indicated in
Fig.~\ref{cfm-2dphase}.  For
$|g_1| > |g_4|$, $p\neq 0,q=0$, so the system is AFM or FM, depending
on the sign of $g_1$.  For $|g_1|<|g_4|$, one finds the columnar dimer
phase (rather than staggered since $g_4<0$).
Note that obtained phase diagram (Fig.~\ref{cfm-2dphase}) is in qualitative agreement
with that of the ladder, which provides {\em a posteriori} support of
our mean-field treatment.  Moreover, it agrees very well with the phase
diagram of the lattice model (\ref{V-full}) obtained in the exact
diagonalization study \cite{sindzingre}.  We should add that DC phase
was not observed in DMRG study \cite{numerics}, presumably due to strong
finite-size effects at small $J_\perp/J$.  Now, since exactly this type
of dimer ordering is known to take place in the spatially isotropic
$J_1-J_2$ model \cite{read-sachdev}, it is natural to conclude that the
DC phase found here extends all way up to $J_\perp=J$.

We thank the Aspen Center for Physics, where this work has began, and
the Kavli Institute for Theoretical Physics, where it continued, for
hospitality.  We are grateful to A. Furusaki and A. Vishwanath for many
stimulating discussions.  O.A.S. thanks A. Abanov, F. Essler, P. Lecheminant and A.
Tsvelik for useful discussions at the Theory Institute of the Brookhaven
National Laboratory, and, especially, A. Nersesyan for illuminating
discussion of the OPE (\ref{ope}).  This research was supported by NSF
Grant No. PHY99-07949 (O.A.S.), DMR-9985255 (L.B.), the Research
Corporation Grant No. CC5491 (O.A.S.) and the Packard Foundation (L.B.).

\end{document}